# Preliminary approaches towards the integration of TSN communications into the NFV architectural framework

Jorge Sasiain, Asier Atutxa, David Franco, Jasone Astorga, Eduardo Jacob

Department of Communications Engineering, University of the Basque Country (UPV/EHU). 48013 Bilbao, Spain.
jorge.sasiain@ehu.eus, asier.atutxa@ehu.eus, david.franco@ehu.eus, jasone.astorga@ehu.eus, eduardo.jacob@ehu.eus

*This paper presents a preliminary architecture for the integration of Time-Sensitive Networking (TSN) communications into the Network Functions Virtualization (NFV) architectural framework. Synergies between functional blocks and constructs of NFV, and components of TSN networks, are investigated in order to arrive at an integrated architecture. Additionally, mechanisms and configuration procedures to enable TSN-compliant, real-time, and virtualized end stations under the NFV framework are explored.*

## I. INTRODUCTION AND MOTIVATION

5G technology is undoubtedly having a significant impact on the industrial manufacturing and automation vertical, being, in fact, a key driver of the fourth industrial revolution (Industry 4.0). Industry 4.0 embraces a series of technologies and paradigms primarily focused around ubiquitous communication scenarios, which enables innovative use cases. In order to materialize all these applications, there is a clear trend towards the virtualization and cloudification of their functionalities [1], encompassing entities and technologies such as Programmable Logic Controllers (PLCs), robotics, IIoT gateways, Cyber-Physical Systems (CPSs), automotive systems, and digital twins. Virtualization increases flexibility and agility, facilitates offloading of complex processing to an edge cloud, and enables coexistence of mixed-criticality applications.

Industrial scenarios often involve closed-loop control systems with stringent requirements of determinism, latency, and reliability that standard Ethernet cannot satisfy. IEEE Time-Sensitive Networking (TSN) standards have been developed to bridge this gap forgoing traditional proprietary fieldbus protocols that lack interoperability. However, challenges arise when trying to incorporate virtualization technologies into TSN, at both performance and configuration management level. The literature primarily focuses on the former aspect, proposing mechanisms for real-time containers and Virtual Machines (VMs) such as real-time task scheduling policies [2], preemptable kernel and co-kernel approaches [3], and specific hypervisor architectures [4]. However, works addressing the orchestration aspect are scarce and focus mainly on specific modules of container platforms like Kubernetes [5].

To the best of our knowledge, initiatives towards a full-fledged integration of time-sensitive communications with virtualization and orchestration architectures is missing. Thus, this paper sketches an initial approach for integrating TSN and ETSI Network Functions Virtualization (NFV), the latter considered a building block of 5G networks [6].

## II. BACKGROUND

IEEE TSN comprises a toolbox of standards which provide functionalities spanning time synchronization, bounded low latency, high reliability, and resource management. A TSN *stream* is an unidirectional flow of data between *end stations*, i.e. from a *Talker* to one or more *Listeners*, that traverses TSN *bridges*, and that belongs to a traffic class identified by the VLAN Priority Code Point (PCP). The 802.1Qcc standard specifies three types of configuration modes for a TSN *domain*. In the fully centralized one, the Centralized Network Configuration (CNC) entity retrieves bridge capabilities and configures them, while the Centralized User Configuration (CUC) entity retrieves application stream requirements and discovers and configures end station capabilities. CUC sends stream requirements to the CNC via the User/Network Interface (UNI). The separation of the network into TSN domains is an administrative decision (e.g. by production lines, machine units, hierarchical network segments, etc.). The primary TSN standard to achieve bounded low latency for periodic traffic is 802.1Qbv, which provides temporal isolation by computing a network-wide schedule where each traffic class is assigned a dedicated time slot enforced by egress device ports. 802.1Qbv is supported by the capabilities provided by the 802.1AS standard to synchronize the clocks of all TSN devices. Multiple TSN domains may share a common working clock domain.

The ETSI NFV architectural framework is concerned with the abstraction, or decoupling, of network functions from physical hardware. The NFV Infrastructure (NFVI) contains and abstracts physical resources for use by Virtual Network Functions (VNFs), which can be composed of multiple sub-components —VMs or containers— to encapsulate a specific networking functionality. VNFs can be chained together to form Network Services (NS). Virtual Links (VLs) interconnect VNF sub-components and VNFs. The Management and Orchestration (MANO) block is subdivided into the Virtualized Infrastructure Manager (VIM), responsible for the management of NFVI resources, the VNF Manager (VNFM), responsible for the lifecycle management of VNFs, and the NFV Orchestrator (NFVO), which offers end-to-end NS orchestration and can interact with OSS systems. Additionally, a WAN Infrastructure Manager (WIM) can provide multi-site connectivity management between NFVI Points of Presence (NFVI-PoPs). NFV and Software-Defined Networking (SDN) are widely regarded as complementary technologies, and NFV standards explicitly address the interworking with SDN.

The remainder of this paper assumes that the fully centralized TSN configuration model is used and that, if communications traverse multiple TSN domains, they share the time synchronization domain. Additionally, the article focuses on enabling support for the 802.1Qbv standard. The proposal is contextualized in the industrial automation and manufacturing vertical under the Industry 4.0 umbrella. In this scenario, time-sensitive applications could be allocated in heterogeneous platforms, such as multi-processor SoCs or FPGAs with embedded Operating Systems that connect to field devices in production lines or robots, as well as to traditional edge cloud resources that provide additional computational capabilities.

## III. ARCHITECTURE-LEVEL INTEGRATION

This section proposes an architecture-level integration between NFV functional blocks and TSN components to enable time-sensitive communications involving applications deployed under the NFV framework using hypervisor-based or container-based virtualization. Two integration modes are considered to address two different communication use cases: (a) integration to enable intra-VIM TSN communications; and (b) integration to enable inter-VIM TSN communications. Use case (a) involves a single TSN domain —mapped 1-to-1 to a NFVI-PoP—, for communications between field devices, local PLCs, local computing, etc., while use case (b) allows to span multiple TSN domains for communications between production lines, to a centralized edge cloud, etc. A high level view of the proposed architecture, which is described in the remainder of this section, is shown in Figure 1.

### A. Intra-VIM Communication Use Case

The intra-VIM communication use case involves the communication between VNFs and their underlying resources located within a single NFVI-PoP. The NFVI incorporates TSN bridges apart from the computing resources. As the VIM is the functional block ultimately responsible for the interaction with the NFVI, the CNC is placed at VIM level, where three implementation possibilities are considered: (a) CNC functionality running in parallel to VIM functionality in a coordinated manner; (b) CNC functionality being implemented as part of the VIM; and (c) CNC functionality being implemented as part of, in parallel to, or on top of an SDN Controller (SDNC) (i.e. as an SDN application) that is located in the VIM. In (c), the CNC can inherit functionalities such as topology discovery from the SDNC.

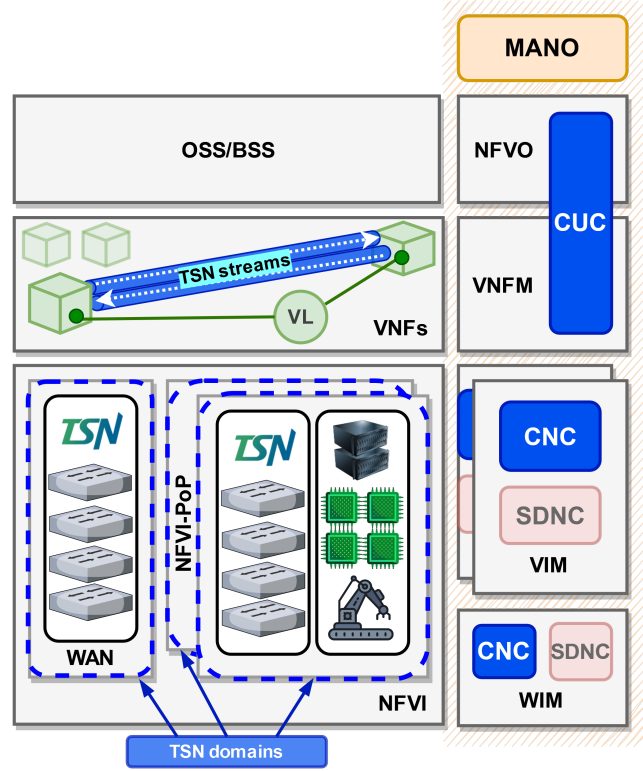

Fig. 1. Proposed architecture for integrating TSN support into the NFV architectural framework. An orange background denotes the NFV MANO functional blocks (right). TSN components and entities are depicted in blue. A simple NS is overlaid in green (left).

The user-facing CUC functionality is provided by the NFVO functional block of NFV MANO. The NFVO receives NS lifecycle requests that, in this case, contain additional configuration that targets the TSN resources. Information that concerns the TSN bridges in the NFVI is sent to the CNC located at VIM level. Hence, information exchanged through the UNI takes place between the NFVO and the VIM, which translates to the Or-Vi reference point in the NFV architectural framework. On the other hand, CUC functionality related to configuring end stations and retrieving their capabilities is implemented in the VNFM via the Ve-Vnfm interface towards the VNFs.

### B. Inter-VIM Communication Use Case

The inter-VIM communication use case involves the communication between VNFs and their underlying resources located in separate NFVI-PoPs. On the one hand, each NFVI-PoP requires the support discussed in the intra-VIM case for the portions of the TSN streams that involve resources in the extent of that NFVI-PoP.

Table I
PROPOSED EQUIVALENCE (BLUE) AND FUNCTIONAL MAPPING (GREEN) BETWEEN TSN AND NFV ENTITIES AND CONSTRUCTS

| TSN | Domain | End station | Stream | CUC | CNC | UNI |
|---|---|---|---|---|---|---|
| NFV | NFVI-PoP or WAN segment | VM, container, or single-component VNF | One direction of VLs or of an NS | NFVO and VNFM | VIM or WIM | Or-Vi or Or-Wi |

On the other hand, additional TSN resources are also found in the WAN or transport network between NFVI-PoPs, which is managed by the WIM. Two WIM types are considered: (a) standalone WIM; and (b) WIM being itself an SDNC or an SDN application running on top of a SDNC. In case (a), CNC functionality can be either embedded in the WIM, or be implemented in parallel or on top of it in a coordinated manner. In case (b), CNC functionality can be implemented as part of, in parallel to, or on top of the SDNC. Like before, the NFVO/VNFM pair is in charge of executing CUC functionality, and the UNI exists between the NFVO and the CNC, which this time is at WIM level. This corresponds to the Or-Wi reference point in the NFV architectural framework.

So far this scenario has assumed that there is a number of TSN domains equal to the number of NFVI-PoPs plus WAN segments, yet there is only a single location housing CUC functionality —the NFVO/VNFM— for all domains. On the one hand, NFV standards do not constrain the management scope of a VIM to one or to all NFVI-PoPs in the NFV architecture. Hence, the considered approach is to have one VIM manage each NFVI-PoP in order to match the scope of each TSN domain. On the other hand, the TSN 802.1Qcc standard specifies each TSN domain to be managed by its own CNC and CUC. However, as the NFVO —and thus CUC— is expected to receive service requests from a centralized OSS, stream requirements do not need to be provided directly by end stations.

Depending on the architecture of the industrial network at hand, it might be possible that no central management segment exists in it to deploy a NFVO/VNFM pair with direct management access to all involved TSN domains. In this case, where each separate MANO controls a subset of all TSN domains, it becomes effectively a concatenation of intra-VIM communications. TSN inter-domain coordination and configuration mechanisms would be required to realize end-to-end streams, but this would be considered outside of the scope of NFV. On the contrary, if the NFVO/VNFM pair has access to management connectivity for all involved TSN domains, a centralized CUC in the NFVO is used, relying on NFV MANO procedures to derive the target VIM —and thus the target CNC— of each service component and involved TSN stream.

### C. Communication with External Systems

It has to be taken into account that streams may not only span end stations that reside inside the NFVI, for example in communications in which industrial sensors, actuators, and/or legacy control systems participate. If the functional lifecycle of such external components can be orchestrated by NFV MANO procedures, they can become part of the NFVI and their behavior modelled as a Physical Network Function (PNF) under MANO's visibility. This would result in a scenario equivalent to the intra-VIM communication one. On the contrary, if the external components cannot be exposed to orchestration by MANO, NFV is not responsible for any resources located outside the NFVI with regards to TSN streams that cross these boundaries.

## IV. PLATFORM AND VNF CONFIGURATION

In addition to the coordinated orchestration of TSN and NFV resources, the other main challenge is that those resources have to satisfy the performance requirements of time-sensitive communications in spite of the virtualization layers. These concerns can be grouped into: minimization of latency and jitter caused by virtualization overhead and resource sharing; time synchronization for VMs/containers; and time-triggered transmission scheduling for VM/container streams. Other requirements are network topology and bridge capabilities discovery, the determination of bridge and propagation delays, and 802.1Qbv schedule synthesis, which are inherent to TSN networks and thus outside of this paper's scope. This section proposes how to manage and carry out the configuration required for TSN support in the NFV environment.

### A. Mechanisms for Time-sensitive Communications with Virtualization

The toolbox of mechanisms that can contribute to providing real-time guarantees and TSN compliance to end stations —with and without virtualization— includes:

- TSN features provided by the Linux ecosystem [7]. This includes Linux PTP for 802.1AS time synchronization in the NIC and system clocks, and Traffic Control Queuing Disciplines (Qdiscs) to implement traffic shaping functionality like 802.1Qbv.
- High-priority, real-time scheduling policies such as SCHED_DEADLINE in Linux, which incorporates awareness of the deadline of the tasks.
- Real-time operating systems (RTOS).
- Real-time co-kernels that run in parallel to the main kernel and are responsible for time-critical workloads.
- The PREEMPT_RT patch for the Linux kernel, which increases preemptability of kernel code.
- Hypervisors with real-time capabilities.
- Hardware resources isolation for VMs/containers, e.g. CPU pinning and SR-IOV passthrough.

The activation of these mechanisms involves hardware and/or software requirements. Notably, hypervisor-based virtualization without resource pinning requires the hypervisor to take charge of VM transmissions according to a global schedule, e.g. dispatching vCPUs appropriately [8].

*B. Stream Requirements Retrieval Procedure*

Whenever a NS lifecycle operation is triggered, the CUC at NFVO level shall provide TSN configuration information to the CNC at VIM or WIM level via the UNI. Multiple CNCs in separate VIMs and/or WIMs may be targeted depending on the placement of the NS components. This information is derived from the user-defined NFV service descriptors, which shall be augmented to accommodate TSN stream requirements. Based on 802.1Qcc specifications, the required information per new stream during an NS instantiation procedure includes:

a. Identification of the end stations and network interfaces participating in the stream.
b. Data frame specification (MAC and IP addresses of participating end stations, and VLAN tag).
c. 802.1Qbv traffic specification, such as period, maximum frame size, frames transmitted in a period, and maximum allowed latency.

The information in c. cannot be derived through native NFV descriptors and procedures. Therefore, additional information shall be provided within the corresponding VL constructs in a VNF or NS descriptor, as appropriate. It has to be taken into account that TSN streams are unidirectional, but VLs are bidirectional. For example, an initial design approach may assume a single Listener per TSN stream and consider a constraint that a NS must map exactly to two TSN streams in the two directions of a communication between two end stations.

*C. Configuration Procedure of End Stations*

After receiving and processing stream requirements, the CNC configures the TSN bridges and sends the result back to the CUC, which concludes by forwarding configuration directives to the end stations. Although several real-time capabilities such as the use of a real-time operating system, kernel, or hypervisor, are platform-wide, specific per-stream VM/container configuration also has to be carried out by the CUC. VNF configuration is conducted by the VNFM via the Ve-Vnfm reference point, so this CUC functionality is allocated to the VNFM functional block. This configuration includes:

- Installation and execution of any required software, such as daemons for 802.1AS time synchronization.
- VLAN configuration of the network interfaces.
- Setting of internal Linux packet priorities (socket option SO_PRIORITY), and mapping to VLAN PCPs.
- Setting of process scheduling policies and priorities.
- Configuration of TSN traffic classes via TSN Qdiscs, such as the Time Aware Priority (TAPRIO) and the Earliest TxTime First (ETF) Qdiscs for 802.1Qbv.

Adhering to the timings specified in the stream requirements is the responsibility of the application running in each VM/container. Because the configuration described above is applied to each VM/container, no specific action is required in NFVI resources during an NS termination procedure (only by the CNC). On the other hand, NS or VNF update actions may involve reconstruction of streams and thus reconfiguration of VMs/containers.

## V. CONCLUSIONS AND OUTLINE

This paper proposes guidelines towards a preliminary approach for supporting TSN in NFV. However, there is considerable room for delving further into integration details and exploring alternative methods and technologies.

*1) Container-based NFV architectures:* following the cloud-native paradigm shift, NFV Rel. 4 has introduced explicit support for containerized applications. Platforms like Kubernetes are gaining traction as orchestration tools. An integration of TSN into Kubernetes could be explored, leveraging its Container Network Interface framework, and Kubernetes Operators to extend its orchestration scope.

*2) Kernel-bypass packet processing technologies:* Many research works point to the use of kernel-bypass and userspace technologies such as XDP and DPDK to improve packet processing performance in the context of TSN, especially when virtualization is involved. Support for the acceleration of VNF packet processing through these technologies could bring performance benefits.

*3) Integration of additional TSN standards:* in addition to 802.1Qbv, the integration with other latency standards like 802.1Qbu and 802.1Qch, and reliability standards like 802.1CB and 802.1Qci, could be considered.

*4) Service Function Chaining:* NFV supports other constructs based on forwarding graphs, where the communication paths between VNFs are fixed, often leveraging SDN. Their mapping to TSN streams could be investigated.

*5) Wireless TSN:* a significant effort is currently being carried out by researchers and SDOs alike to enable wireless TSN communications, e.g. in 3GPP Release 16 onwards. 5G-TSN integration is a research line of its own; still, significant complementary synergies could be found.

## ACKNOWLEDGEMENTS

This work was supported in part by the Spanish Ministry of Science and Innovation through the national project (PID2019-108713RB-C54) titled "Towards zeRo toUch nEtwork and services for beyond 5G" (TRUE-5G), and in part by the Basque Government through the project Social Network for Machines (SONETO) (KK-2023/00038) of the ELKARTEK Program.